\documentclass{epl}
\usepackage{amssymb,graphicx,psfrag,amsmath}
\title{Out of plane optical conductivity in d-wave superconductors}
\shorttitle{Out of plane optical conductivity in d-wave \dots}
\author{Bal\'azs D\'ora\inst{1} \and Kazumi Maki\inst{2} 
\and Attila Virosztek\inst{1,3}}
\institute{
\inst{1} Department of Physics, Technical University of Budapest, H-1521 Budapest, Hungary \\
\inst{2} Department of Physics and Astronomy, University of Southern California, Los Angeles
CA 90089-0484, USA \\
\inst{3} Research Institute for Solid State Physics and Optics, P.O.Box 49,
H-1525 Budapest, Hungary}
\shortauthor{Bal\'azs D\'ora \etal}


\date{}

\begin{document}

\maketitle

\begin{abstract}
We study theoretically the out of plane optical conductivity of d-wave
superconductors in the presence of impurities at $T=0$K. Unlike the usual approach,
we assume that the interlayer quasi-particle transport is due to coherent
tunneling. The present model describes the $T^2$ dependence of the out of
plane superfluid density observed in $YBCO$ and $Tl2201$ for example. In the
optical conductivity there is no Drude peak in agreement with experiment, and
the interlayer Josephson tunneling is also assured in this model. In the unitary
limit we predict a step like behaviour around $\omega\sim\Delta$ in both the real
and imaginary part of the optical conductivity.

\end{abstract}

\section{Introduction}

It is now well established that both the thermodynamics and transport
properties of high $T_c$ cuprate superconductors are well described by the
BCS theory of d-wave superconductors \cite{maki1,maki2,dsc1,dsc2}. More recently it is shown that the
electron-doped high $T_c$ superconductors also belong to d-wave \cite{dsc3,dsc4,dsc5}.
Also after a long controversy \cite{et2-1}, d-wave superconductivity in $\kappa-(ET)_2$
salts is emerging \cite{et2-2,et2-3}. Therefore single crystals of $YBCO$, $Bi2212$ and
$\kappa-(ET)_2$ salts will provide the excellent testing ground of d-wave
superconductors. In particular the planar transport in high $T_c$ cuprate
superconductors is well described in terms of the BCS like quasi-particles \cite{maki3,hirsch1,hirsch2,maki4,graf,moi}.
For example the $T$-linear dependence of the in plane superfluid density is
established for the first time in $YBCO$ \cite{ybco1}. 
In contrast the out of plane transport in high $T_c$ cuprates is rather
controversial. First of all the absence of the Drude peak in the optical
conductivity of $Bi2212$ and $YBCO$ is very difficult to interpret in
terms of the usual quasi-particle transport \cite{et2-3,ybco2}. As an alternative, many
authors consider the incoherent tunneling \cite{graf,incoh}. In this way it is easy to obtain
a flat optical conductivity as $\omega$ goes to zero. On the other hand
there will be no interlayer Josephson coupling \cite{et2-3}, contrary to the observation
of the Josephson coupling seen experimentally in $Bi2212$ \cite{josephson1}. Further
Josephson plasmons are observed in $Bi2212$, $YBCO$ \cite{josephson2} and
in $\kappa-(ET)_2$ salts \cite{josephson3}.

Therefore in the following we assume that the interlayer transport is
described in terms of coherent tunneling \cite{et2-3,coh}. Let us consider the
quasi-particle transport from one layer to another. If it is usual
quasi-particle transport $\bf k=k^\prime$, where $\bf k$ and $\bf k^\prime$
are the planar quasi-particle momentum in these two layers. In this model
the out of plane transport is essentially the same as the in plane
transport. So there should be a Drude tail in the optical conductivity. In
the incoherent tunneling model $\bf k\ne k^\prime$; $\bf k$ and $\bf
k^\prime$ are uncorrelated. So there will be no Josephson coupling between
two layers contrary to the experiment. In the coherent tunneling model, we
assume $\bf k\ne k^\prime$ but $\bf k \parallel k^\prime$, or at least some
directional correlation. This is the model considered by Ambegaokar and
Baratoff \cite{baratoff1,baratoff2} and they call this "specular transmission". We believe that this
is the only model which gives the interlayer Josephson coupling in
unconventional superconductors.

More generally some mixture of the coherent and the incoherent tunneling is
possible. But we limit ourselves to the simplest case.

The first important consequence of this choice is the temperature
dependence of the out of plane superfluid density which is different from
the in plane one. In particular both in the clean limit and in the presence
of impurities the out of plane superfluid density exhibits the $T^2$
dependence as seen in $YBCO$ \cite{T21}, $Tl2201$ \cite{T22} and
$\kappa-(ET)_2$ salts \cite{et2-3}.

\section{Superfluid density at $T=0$K}

\begin{figure}[h]
\psfrag{x}[t][b][1][0]{$\Gamma/\Gamma_c$}
\psfrag{y}[b][t][1][0]{$\rho_{s \perp}(\Gamma,0)$} 
\psfrag{z}[b][t][1][0]{$\rho_{s,ab}(\Gamma,0)$}
\twofigures[width=7cm,height=7cm]{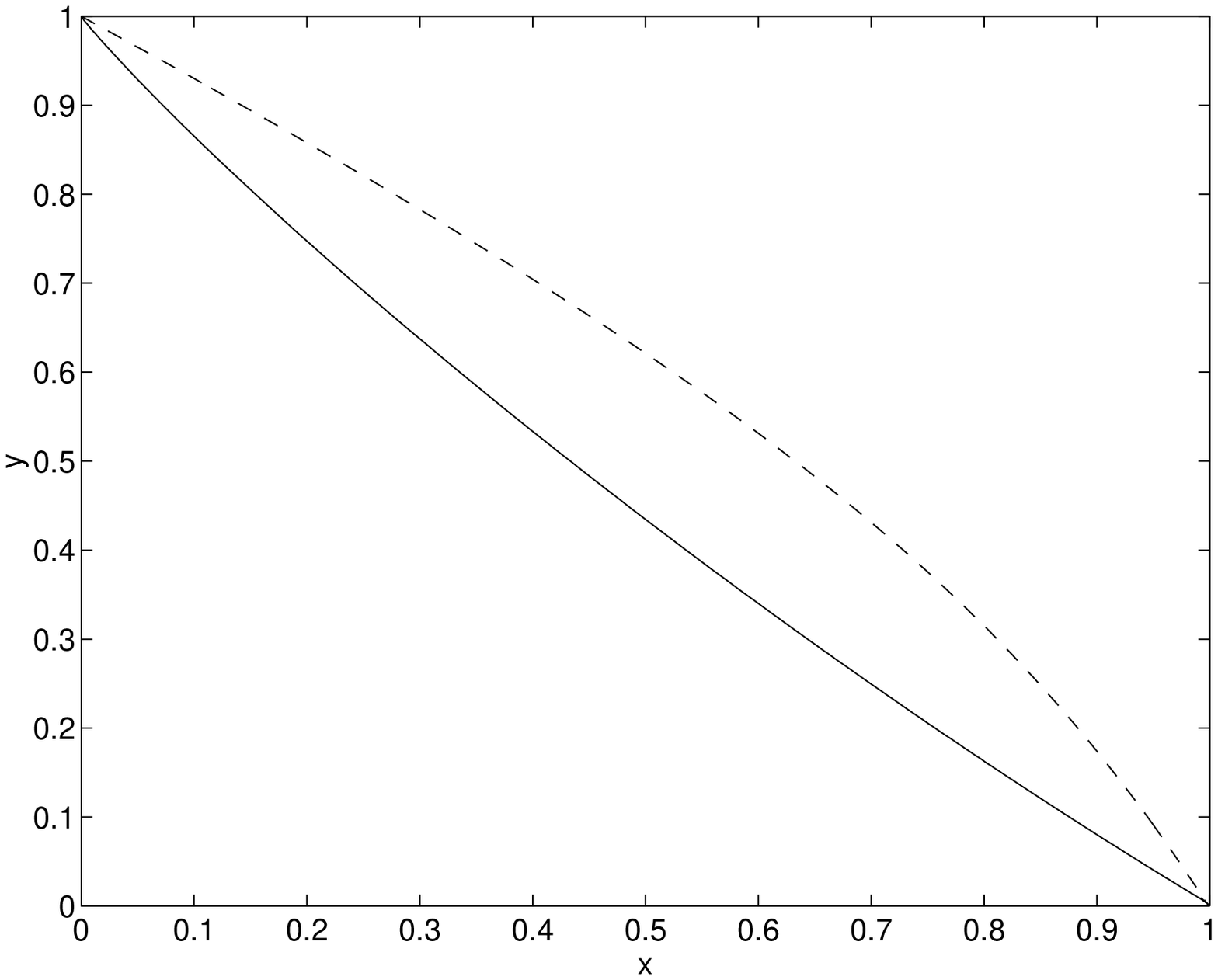}{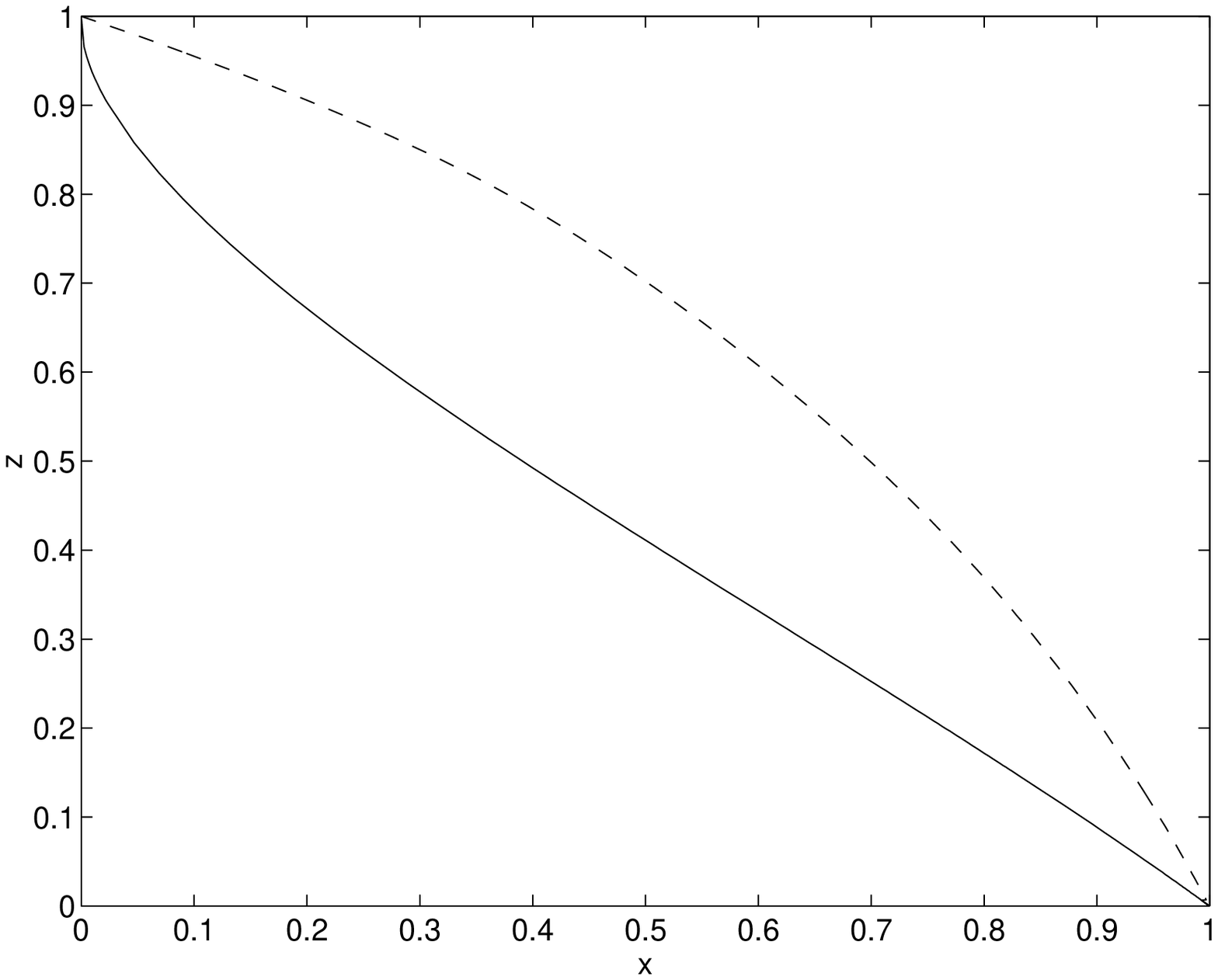}

\caption{The out of plane superfluid density is plotted as a function of
$\Gamma/\Gamma_c$ in the unitary (solid line) and Born (dashed line) limit.}
\label{fig:superfd}
\caption{The in plane superfluid density is plotted as a function of
$\Gamma/\Gamma_c$ in the unitary (solid line) and Born (dashed line) limit.}
\label{fig:absuperfd}
\end{figure}

As in \cite{moi} we consider d-wave superconductivity in the presence of
impurities. The effect of impurity scattering is incorporated by renormalizing the frequency
 in the quasi-particle Green's function \cite{hirsch1,hirsch2,maki4,graf,moi}:
\begin{eqnarray}
\frac{\omega}{\Delta}=u+\alpha\frac\pi 2
\frac{\sqrt{1-u^2}}{uK\left(\frac{1}{\sqrt{1-u^2}}\right)}, \\
\frac{\omega}{\Delta}=u-\alpha\frac2\pi
\frac{u}{\sqrt{1-u^2}}K\left(\frac{1}{\sqrt{1-u^2}}\right)
\end{eqnarray}
for the unitary and the Born limit, respectively, where
$u=\tilde{\omega}/\Delta$, $\alpha=\Gamma/\Delta$, $K(z)$ is the complete
elliptic integral of the first kind, $\tilde{\omega}$ is the renormalized
frequency and $\Gamma$ is the scattering rate. We have summarized some
properties of these systems in \cite{moi}.
Then the superfluid density at $T=0$K is obtained from the
integral:
\begin{eqnarray}
I_s=2\pi T\sum_n\left\langle\frac{\Delta^2 f^2}{\tilde{\omega}_n^2+\Delta^2
f^2}\right\rangle=\Delta\int\limits_0^\infty
dx\left(1-\frac{u}{\sqrt{1+u^2}}\right)= \nonumber \\
=\Delta\left(\frac{1}{\sqrt{1+C_0^2}}-\frac\pi 2 \alpha\int\limits_{C_0}^\infty
\frac{du}{u(1+u^2)}\left[K\left(\frac{1}{\sqrt{1+u^2}}\right)\right]^{-1}\right)\\
=\Delta\left(\frac{1}{\sqrt{1+C_0^2}}-\frac 2 \pi \alpha\int\limits_{C_0}^\infty
\frac{du u}{(1+u^2)^2}K\left(\frac{1}{\sqrt{1+u^2}}\right)\right)
\end{eqnarray}
for the unitary and the Born limit, respectively, where $f=\cos(2\phi)$,
$\langle \dots \rangle$ means average of $\phi$ and $C_0$ is given
by 
\begin{equation}
C_0^2=\frac \pi 2 \alpha\sqrt{1+C_0^2}\left[
K\left(\frac{1}{\sqrt{1+C_0^2}}\right)\right]^{-1}\label{c0unit}
\end{equation}
and 
\begin{equation}
\sqrt{1+C_0^2}=\frac 2\pi \alpha
K\left(\frac{1}{\sqrt{1+C_0^2}}\right)\label{c0Born}
\end{equation}
for the unitary limit and the Born limit, respectively. The out of plane
superfluid density is obtained from 
\begin{equation}
\rho_{s \perp}(\Gamma,0)=I_s/\Delta_{00},
\end{equation}
where $\Delta_{00}$ is the superconducting order parameter at $T=0$K and
$\Gamma=0$. The superfluid density at $T=0$K in the presence of impurities
is shown versus $\Gamma/\Gamma_c$ in fig. \ref{fig:superfd}, where
$\Gamma_c=0.8819T_{c0}$. In the unitary limit the decrease of $\rho_{s
\perp}(\Gamma,0)$ is much steeper, because stronger impurity effect destroys
superconductivity with more efficacity. At small temperatures $\rho_{s,\perp}$ 
decreases with $T^2$ as well as the penetration depth \cite{et2-3}, contrary to 
the linear change of the in plane superfluid density \cite{maki3}. As a comparison,
we show also the in plane superfluid density at $T=0$K versus $\Gamma/\Gamma_c$ in
fig. \ref{fig:absuperfd}. It is obtained as
\begin{gather}
\rho_{s,ab}=2\pi T\sum_{n=0}^\infty \left\langle\frac{\Delta^2 f^2}{(\tilde{\omega}_n^2
+\Delta^2f^2)^{\frac32}}\right\rangle= \nonumber \\
=1-\frac{\alpha}{C_0}+
\alpha\int\limits_{C_0}^\infty\frac{du}{u^2}\left[1-\frac{E\left(\frac{1}{\sqrt{1+u^2}}\right)}
{K\left(\frac{1}{\sqrt{1+u^2}}\right)}\right]^2
\\
=1-\frac{C_0}{\alpha}-
\frac{4}{\pi^2}\alpha\int\limits_{C_0}^\infty\frac{du}{1+u^2}
\left[K\left(\frac{1}{\sqrt{1+u^2}}\right)-E\left(\frac{1}{\sqrt{1+u^2}}\right)
\right]^2
\end{gather}
in the unitary and Born limit, respectively. Here the two limits are farther from
each other.

\section{Optical conductivity at $T=0$K}

The out of plane optical conductivity is given by
$\sigma(\omega)=\sigma_1(\omega)+i\sigma_2(\omega)$, where
\begin{gather}
\sigma_1(\omega)=-\frac{\sigma_n}{2\omega} \textmd{Re}I(\omega),\label{resigma}\\
\omega\sigma_2(\omega)=-\frac{\sigma_n}{2} \textmd{Im}\left(I(\omega)+2
\int_{-\infty}^\infty\frac{1}{e^{\beta x}+1}F(u(x),u(x-\omega))dx\right),\label{imsigma}
\end{gather}
where 
\begin{eqnarray}
I(\omega)=\int_{-\infty}^\infty \frac 1 2 \left(\tanh\left(\frac{\beta x}{2}\right)-\tanh\left(
\frac{\beta(x+\omega)}{2}\right)\right)\times \nonumber \\
\times(F(u(x+\omega),\overline{u}(x))-F(u(x+\omega),u(x)))
dx \label{integral}
\end{eqnarray}
and
\begin{equation}
F(u,u^\prime)=\left\langle
\frac{uu^\prime+f^2}{\sqrt{f^2-u^2}\sqrt{f^2-{u^\prime}^2}}\right\rangle.\label{fuggveny}
\end{equation}
Note that the above expression would be the same, if we use the
Mattis-Bardeen approximation \cite{mattis} commonly used for s-wave superconductors. But
we emphasize that we don't assume $\alpha\gg 1$ which would be disastrous
for unconventional superconductors but we obtain eqs. \ref{resigma}, 
\ref{imsigma}, \ref{integral}, \ref{fuggveny} from the coherent
tunneling model, since all high $T_c$ cuprates are inherently in the clean
limit (i.e. $\alpha\ll 1$). Also eq. \ref{resigma} in the
$\alpha\rightarrow 0$ limit agrees with $\sigma_1(\omega)$ given in \cite{maki3}.
 Also now it is clear why $\sigma_1(\omega)$ in \cite{maki3} describes the 
frequency dependence of $\sigma_1(\omega)$ determined from the transmission
experiment on $Bi2212$ thin film \cite{Bi2212}.

\begin{figure}[h]
\psfrag{x}[t][b][1][0]{$\Gamma/\Gamma_c$}
\psfrag{y}[b][t][1][0]{$\sigma_1(0)/\sigma_n$}
\onefigure[width=7cm,height=7cm]{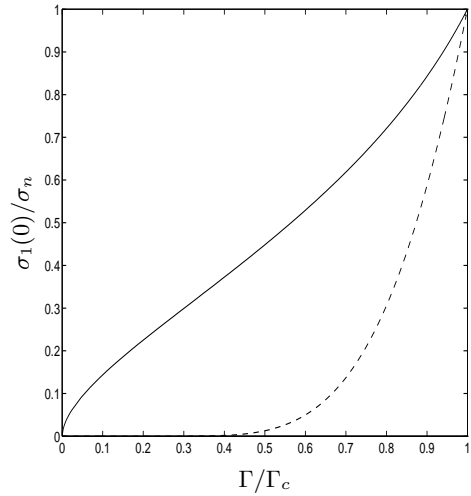}

\caption{The dc conductivity is plotted as a function of
$\Gamma/\Gamma_c$ in the unitary (solid line) and Born (dashed line) limit.}
\label{fig:dc}
\end{figure}

We show in fig. \ref{fig:dc} the dc limit of $\sigma_1(\omega)$ versus
$\Gamma/\Gamma_c$ for the unitary and the Born limit. In terms of $C_0$ it
is given as
\begin{equation}
\frac{\sigma_1(0)}{\sigma_n}=\frac{C_0}{\sqrt{1+C_0^2}},
\end{equation}
where $C_0$ has been defined in eqs. \ref{c0unit}, \ref{c0Born}. In particular in
the limit of $\alpha\ll 1$, $C_0$ is given by
\begin{eqnarray}
C_0\cong \left(\frac \pi 2 \alpha
\left[\ln\left(4\sqrt{\frac{2}{\pi\alpha}}\right)\right]^{-1}\right)^{\frac
1 2}\\
C_0\cong 4 \exp\left(-\frac{\pi}{2\alpha}\right)
\end{eqnarray}
for the unitary and Born limit, respectively. Therefore $\sigma_1(0)$ in
the Born limit is practically zero for $\Gamma/\Gamma_c\le 0.4$.

\begin{figure}
\psfrag{x}[t][b][1][0]{$\omega/\Delta_{00}$}
\psfrag{y}[b][t][1][0]{$\sigma_1(\omega)/\sigma_n$}
\twofigures[width=7cm,height=7cm]{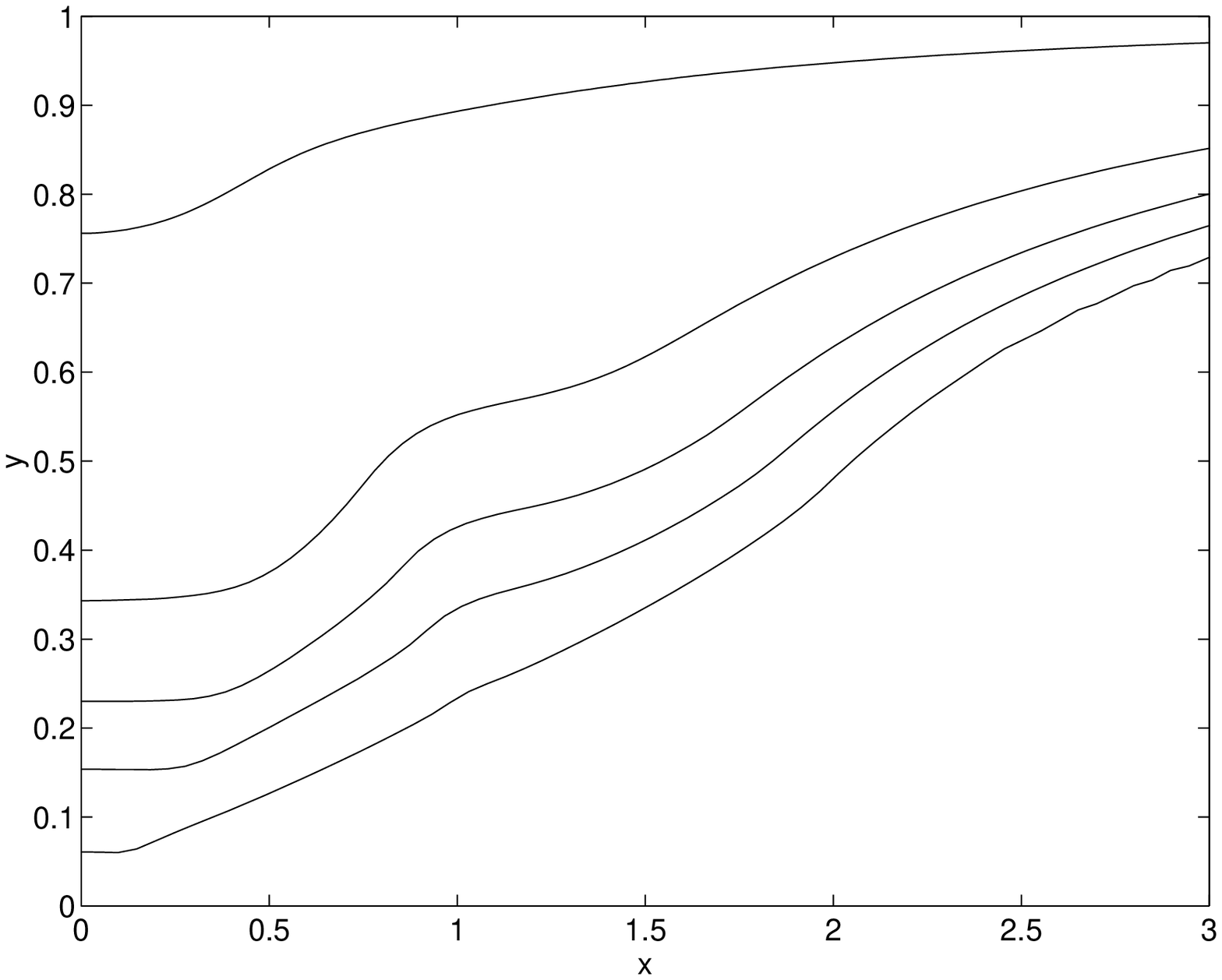}{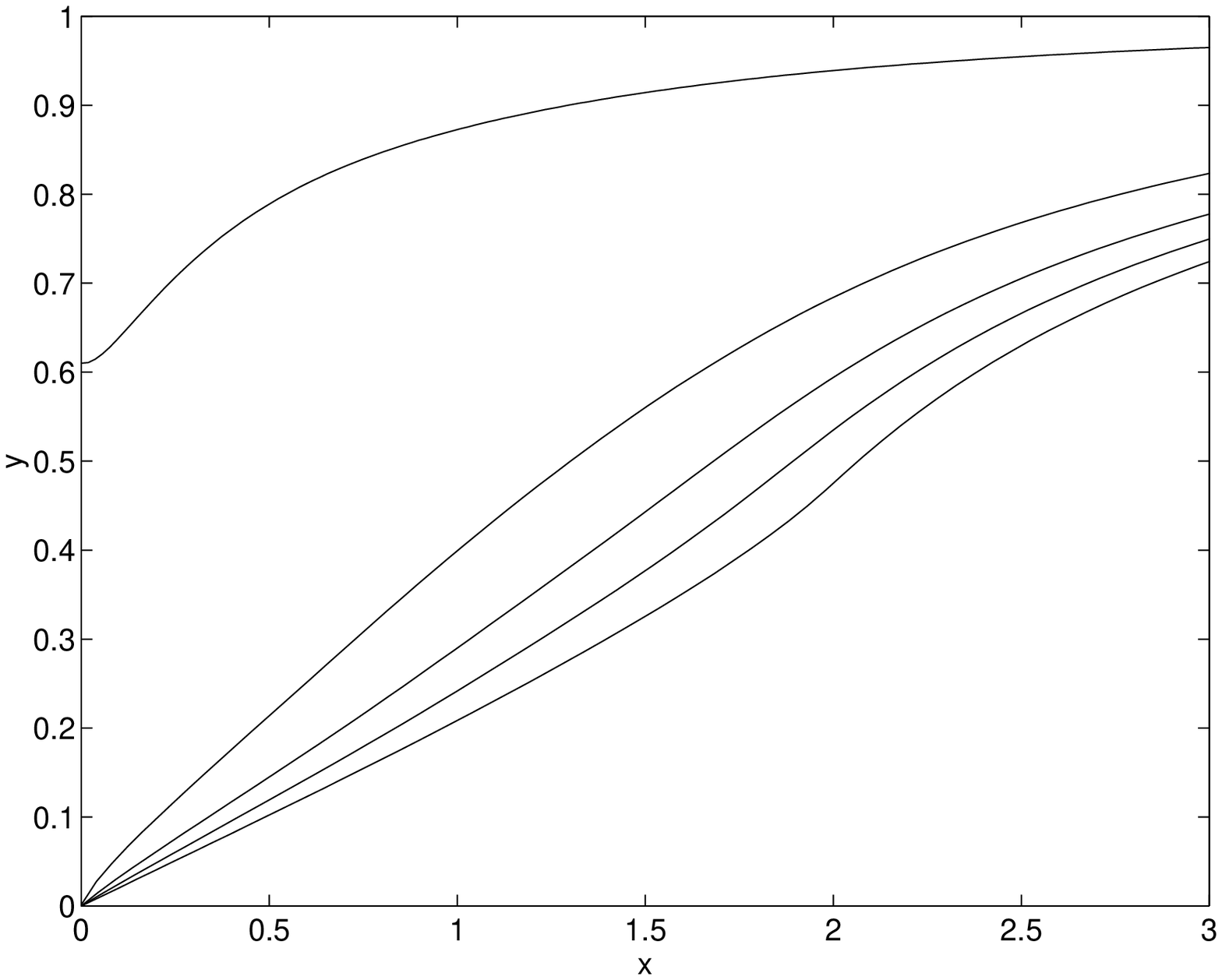}

\caption{The real part of the
optical conductivity is shown in the unitary limit for $\alpha$=0.01,
0.05, 0.1, 0.2 and 1 from bottom to top.}
\label{fig:revezu}
\caption{The real part of the
optical conductivity is shown in the Born limit for $\alpha$=0.01,
0.05, 0.1, 0.2 and 1 from bottom to top.}
\label{fig:revezb}
\end{figure}

We show in fig. \ref{fig:revezu} and \ref{fig:revezb} $\sigma_1(\omega)$ versus $\omega/\Delta_{00}$ for the unitary and the Born limit, respectively. In the unitary limit
$\sigma_1(\omega)$ rises with a step like feature around
$\omega=\Delta$, while in the Born limit it increases monotonically with
$\omega$.  

\begin{figure}
\psfrag{x}[t][b][1][0]{$\omega/\Delta_{00}$}
\psfrag{y}[b][t][1][0]{$\omega\sigma_2(\omega)/\Delta_{00}\sigma_n$}
\twofigures[width=7cm,height=7cm]{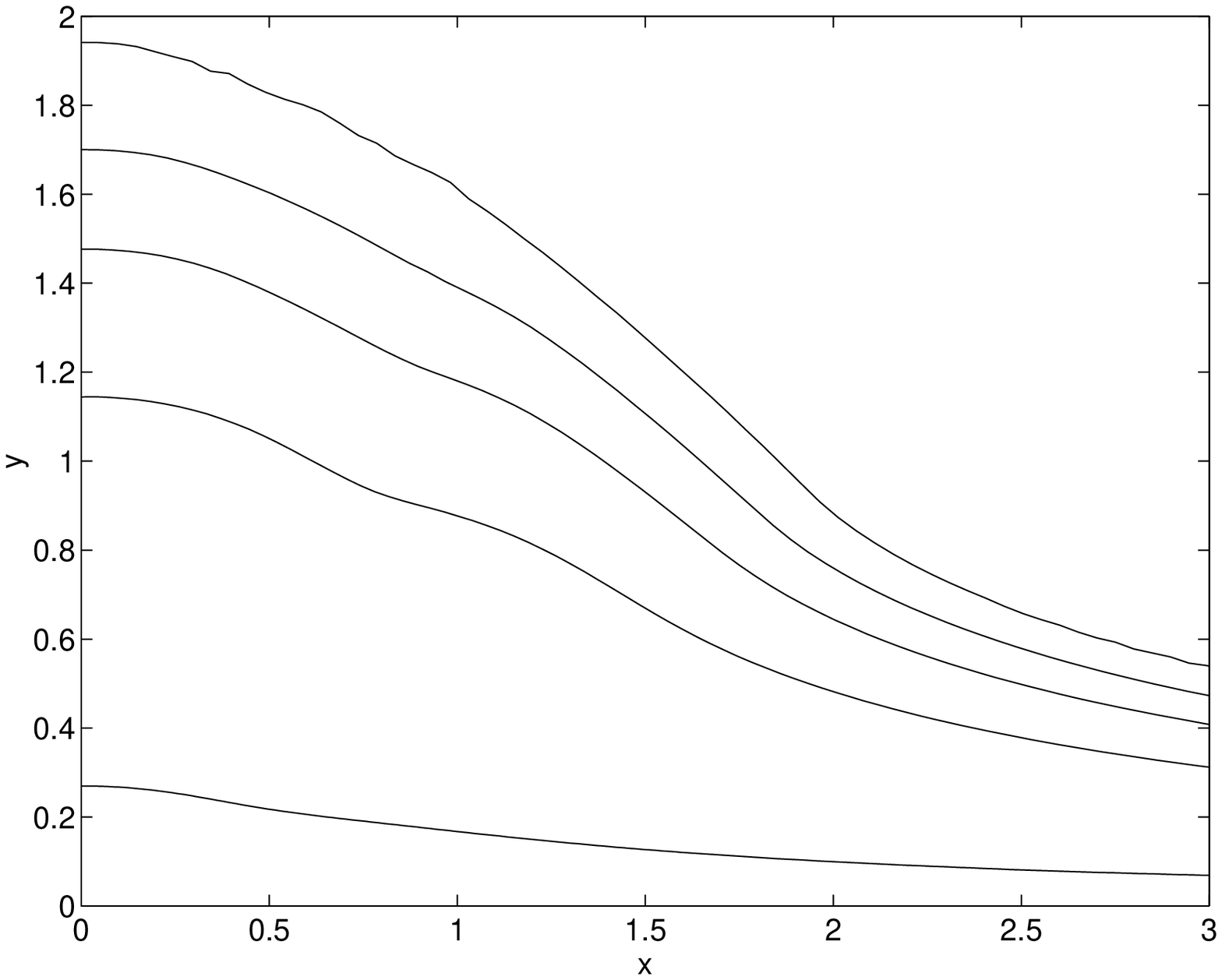}{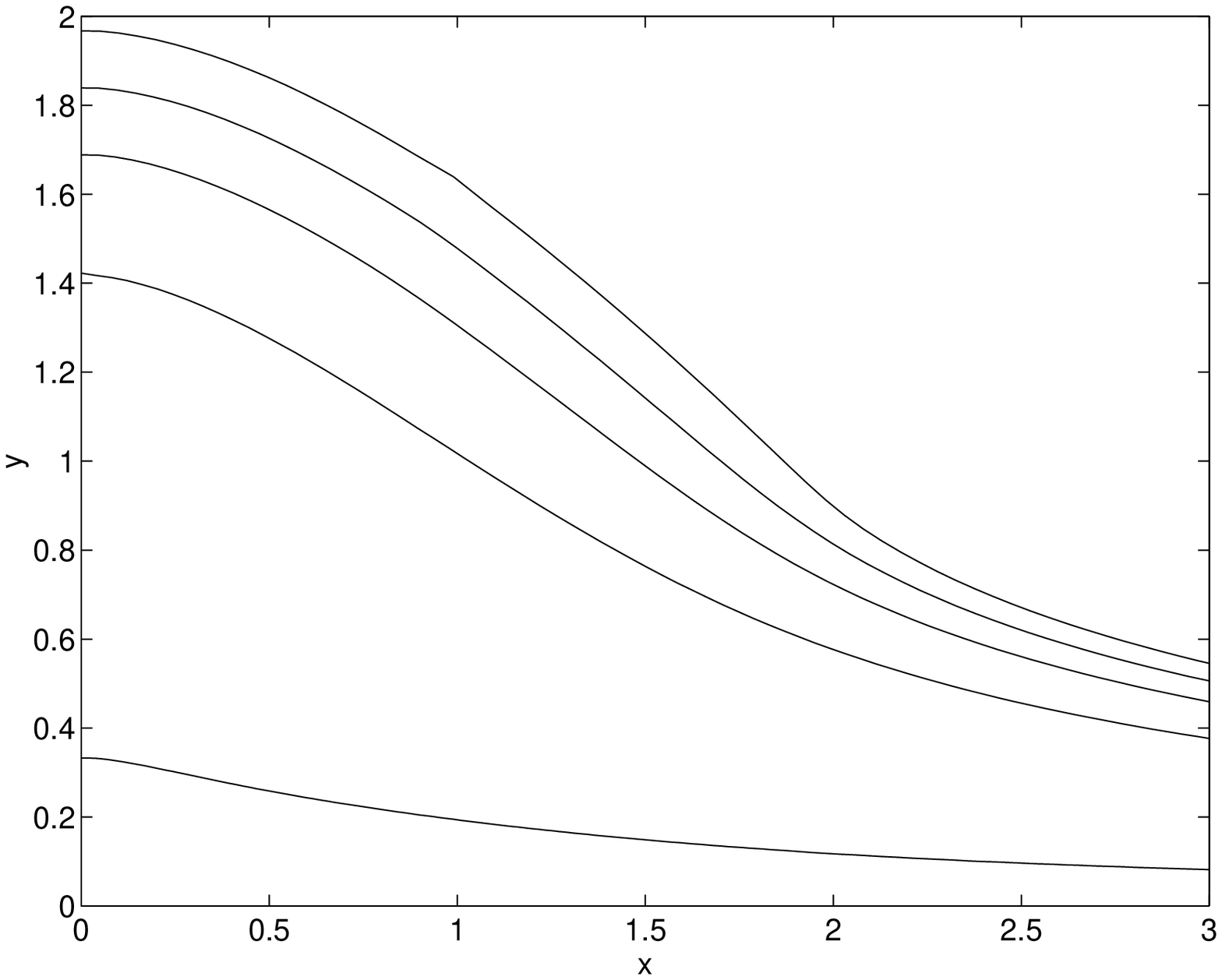}

\caption{$\omega\sigma_2(\omega)$ is shown in the unitary limit for $\alpha$=0.01,
0.05, 0.1, 0.2 and 1 from top to bottom.}
\label{fig:imvezu}
\caption{$\omega\sigma_2(\omega)$ is shown in the Born limit for $\alpha$=0.01,
0.05, 0.1, 0.2 and 1 from top to bottom.}
\label{fig:imvezb}
\end{figure}

$\omega\sigma_2(\omega)$ is shown in fig. \ref{fig:imvezu} and \ref{fig:imvezb} as a function of
$\omega/\Delta_{00}$ for the unitary and the Born limit. Also
$\omega\sigma_2(\omega)$ approaches zero like $\omega^{-1}$ as $\omega$
goes to infinity. So in this sense the Tinkham-Ferrell sum rule should be
obeyed in the present model.
Of course $\omega\sigma_2(\omega)$ decays only slowly with
$\omega$. Therefore the energy cut off may be the problem. Also in the
$\omega\rightarrow 0$ limit, we obtain $\omega\sigma_2(\omega)=2\sigma_n
\Delta_{00}\rho_{s\perp}$ as for the imaginary part of the in
plane conductivity.

\section{Concluding remarks}

We have calculated for the first time the out of plane optical conductivity
in d-wave superconductors at $T=0$K within the coherent tunneling
model. This simple model can describe both the absence of the Drude tail
and the presence of the interlayer Josephson coupling. On the other hand
the present model cannot describe the deviation from the Ferrell-Tinkham
sum rule reported in $Bi2212$ \cite{TFBi2212}. Therefore a further refinement of the
present model may be required. In a future paper we explore the temperature
dependence of the out of plane optical conductivity.

\section{Acknowledgments}
We are
benefited from useful discussions with and help of Hyekyung Won. One of
the authors (B. D.) gratefully acknowledges the hospitality at the University
of Southern California, Los Angeles, where part of this work was done. This work
was supported by the Hungarian National Research Fund under grant numbers
OTKA T032162 and T029877, and by the Ministry of Education under grant number
FKFP 0029/1999.

\bibliographystyle{unsrt}
\bibliography{caxis}
\nocite{*}

\end{document}